\documentclass[prb,twocolumn,showpacs]{revtex4}

\usepackage{graphicx}
\usepackage{amsmath}
\usepackage{amsfonts}
\usepackage{bm}
\usepackage{bbm}

\begin{document}

\title{Polariton amplification in a multiple-quantum-well resonant photonic crystal}

\author{S. Schumacher}
\author{N. H. Kwong}
\author{R. Binder}

\affiliation{College of Optical Sciences,
             University of Arizona,
             Tucson, Arizona 85721, USA}

\date{\today}


\begin{abstract}
Based on a microscopic many-particle theory we study the
amplification of polaritons in a multiple-quantum-well resonant
photonic crystal. For the Bragg-spaced multiple quantum wells under
investigation we predict that in a typical pump-probe setup
four-wave mixing processes can lead to an unstable energy transfer
from the pump into the probe and the background-free four-wave
mixing directions. We find that under certain excitation conditions
this phase-conjugate oscillation induced instability can lead to a
large amplification of the weak probe pulse.
\end{abstract}

\maketitle

\textsc{Introduction} -- In recent years polariton amplification has
been extensively studied for semiconductor quantum wells (QWs)
embedded inside a planar
microcavity.\cite{Savvidis2000,Huang2000,Ciuti2000,Stevenson2000,Saba2001,Whittaker2001,Ciuti2003,Baumberg2005,Keeling2007}
In this Letter we address the question whether similar amplification
can happen in Bragg-spaced multiple quantum wells (BSQWs) not
embedded in a microcavity. In the conventional case of QWs in planar
microcavities the coupling of the excitonic resonances of the QW to
the optical modes confined inside the cavity leads to the formation
of an in-plane (parallel to the QW plane) dispersion with lower and
upper polariton branches (LPB and UPB, respectively). In the strong
coupling regime this dispersion strongly differs from the bare
in-plane cavity and exciton dispersions. The four-wave mixing (FWM)
processes driving the amplification process in this system strongly
benefit from the specific shape of the LPB that for a specific pump
angle allows for triply-resonant\cite{Savvidis2000,Ciuti2000}
(phase-matched) scattering of pump-excited polaritons into pairs of
polaritons in the probe and background-free FWM (BF-FWM) directions.
However, also important for the polariton amplification in planar
microcavities, the strong exciton-cavity mode coupling suppresses
pump-induced excitation-induced dephasing (EID) from two-exciton
correlations for excitation on the
LPB.\cite{Kwong2001a,Kwong2001b,Savasta2003} Then, the
phase-conjugate oscillation (PCO) induced gain can overcome
intrinsic and EID losses to the coherently driven polarizations.
This in turn brings the system above the unstable amplification
threshold and enables strong amplification to occur.

\begin{figure}[b]
\hspace*{-1.5mm}\includegraphics[scale=0.285]{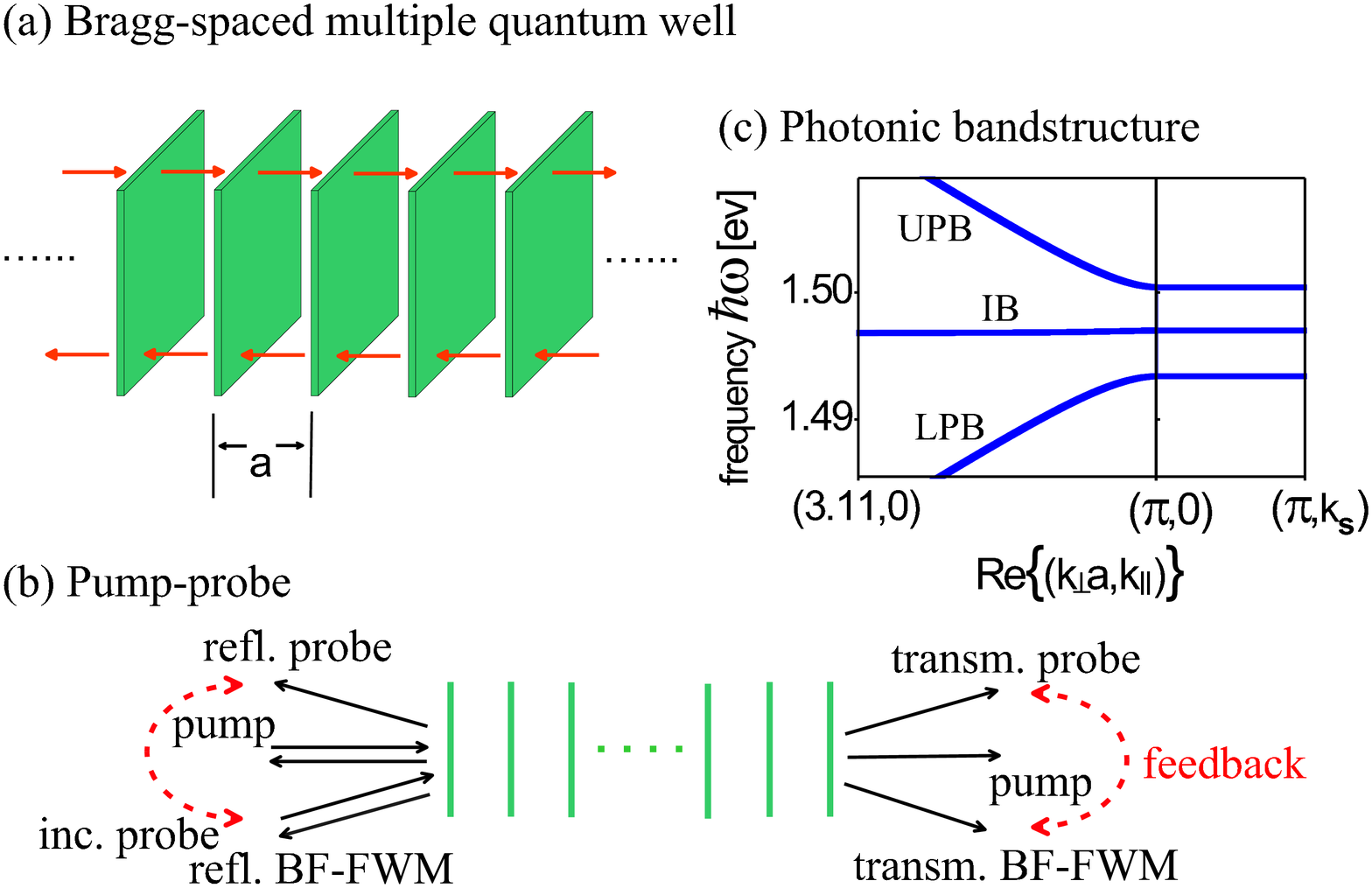}
\caption{\label{Fig1schematics}(color online) (a) Illustration of
light propagation inside a periodic multiple quantum-well structure,
including propagating and counter-propagating field components. (b)
Schematic of a pump-probe geometry, including pump, probe, and
background-free four-wave mixing (BF-FWM) signals. The fundamental
phase-conjugate oscillation feedback between probe and BF-FWM
direction leading to the polariton amplification is indicated. (c)
Photonic bandstructure for an infinite Bragg-spaced multiple
quantum-well structure, showing the intermediate band (IB), the
stop-band regions where no light propagation inside the structure is
possible, and the two polariton bands energetically above (UPB) and
below (LPB) the stop-bands. The dependence of the bandstructure on
the in-plane momentum $k_\parallel$ is neglected for the considered
small angles of incidence related to the in-plane momentum $k_s$.}
\end{figure}

Whereas in the polariton amplification in planar microcavities the
propagation parallel to the plane of the embedded QW is most
important, in multiple QW structures the radiative coupling of the
QWs also leads to polariton effects for propagation perpendicular to
the planes of the QWs.\cite{Deych2000,Ikawa2002,Stroucken1996}
However, in an arbitrarily-spaced (periodic) multiple QW structure
[cf. Fig.~\ref{Fig1schematics}(a)] the required conditions for an
efficient amplification of polaritons will most likely never be met:
Although a degenerate configuration -- where probe and BF-FWM
signals play equal roles in the amplification process [cf.
Fig.~\ref{Fig1schematics}(b)] -- always allows for doubly-resonant
FWM, for pump excitation close to or above the exciton resonance of
the QWs additional EID from two-exciton correlations would inhibit
any observation of amplification.\cite{Schumacher2006}

This situation drastically changes by choosing a BSQW
structure,\cite{Huebner1996,Scharschmidt2004,Nielsen2004,Yang2005}
namely a multiple QW structure with inter-QW distances close to half
the optical wavelength of the QW excitons. Then a one-dimensional
resonant photonic crystal is formed with an effective polariton
dispersion that is strongly modified: A sufficiently long BSQW
structure exhibits an almost perfect stop-band above and below the
exciton resonance of the single QWs where no propagating modes
exist, cf. Fig.~\ref{Fig1schematics}(c). However, more importantly
for this work, the system exhibits effective polariton resonances at
the lower and upper edges of these stop-bands where the polariton
dispersion is flat.\cite{Yang2005}

In this Letter we show how the polariton dispersion of a BSQW
structure can lead to strong amplification of polaritons in a
typical pump-probe setup. Under certain excitation conditions large
time-integrated probe and BF-FWM gain and corresponding (almost)
exponential signal growth over time is found in our numerical
simulations.

\textsc{Microscopic Theory} -- Our theoretical analysis is based on
a combination of a microscopic many-particle theory for the coherent
optical QW response and a time-dependent transfer matrix approach to
properly include the propagation of the optical fields inside the QW
structure. The full numerical analysis is supported by a formal
steady-state linear stability analysis focusing on those
polariton-polariton interactions dominant in the amplification
process. The many-particle part of our theory is formulated in the
framework of the dynamics-controlled truncation
approach\cite{Axt1994a,Oestreich1995} and includes all coherent
optically-induced third order nonlinearities, i.e.,
phase-space-filling, excitonic mean-field (Hartree-Fock) Coulomb
interaction and two-exciton correlations on a microscopic level. We
use a two-band model (including spin-degenerate conduction and
heavy-hole valence band) with the usual circular dipole selection
rules for quasi-normal incidence for the optically induced interband
transitions in the GaAs QWs.\cite{BSInstfootnote1} Restricting
ourselves to excitation spectrally below the bare exciton resonance,
we account for the dominant contributions to the coherent optical
response of the QWs by evaluating the polarization equations in the
1s heavy-hole exciton approximation.
\cite{Donovan2001,Takayama2002,Buck2004,Schumacher2005a} The
propagation of the light fields in the QW structure is included via
a time-dependent transfer matrix approach that in the linear regime
for an infinite BSQW structure gives the polariton bandstructure
shown in Fig.~\ref{Fig1schematics}(c).\cite{Yang2005} To simulate a
typical pump-probe setup as illustrated in
Fig.~\ref{Fig1schematics}(b) we apply an in-plane momentum
decomposition of field and polarization: the pump has zero in-plane
momentum $k=0$ and we chose the finite (albeit small) in-plane
momenta $k_s$ and $-k_s$ for probe and BF-FWM, respectively. The
in-plane dispersion of the excitons is neglected along with possible
small changes to the polariton dispersion for probe and BF-FWM
polaritons not strictly propagating in normal incidence but carrying
the small in-plane momenta $k_s$ and $-k_s$, respectively. We go
beyond an evaluation of the theory on a strict $\chi^{(3)}$ level by
self-consistently calculating the resulting exciton and two-exciton
polarization dynamics up to arbitrary order in the optical field
(the equations of motion are linearized in the weak probe
field).\cite{Oestreich1999,Buck2004,Schumacher2006} Correlations
involving more than two excitons and those involving incoherent
excitons are neglected. These effects are not expected to
qualitatively alter the presented results for the considered
coherent exciton densities of $\sim10^{10}\,\mathrm{cm^{-2}}$ and
for excitation well below the exciton resonance. Via this
self-consistent solution the coupling of the probe signal to the
BF-FWM signal is included which provides the basic PCO FWM feedback
mechanism that can lead to instability as we discuss in more detail
in Ref.~\onlinecite{Schumacher2006}. In the present study we
concentrate on co-circular ($\sigma_+\sigma_+$) pump-probe
excitation and keep the discussion of polarization dependencies from
spin-dependent polariton scattering\cite{Nguyen2007,Schumacher2007a}
for the future.

\begin{figure}[t]
\includegraphics[scale=1.48]{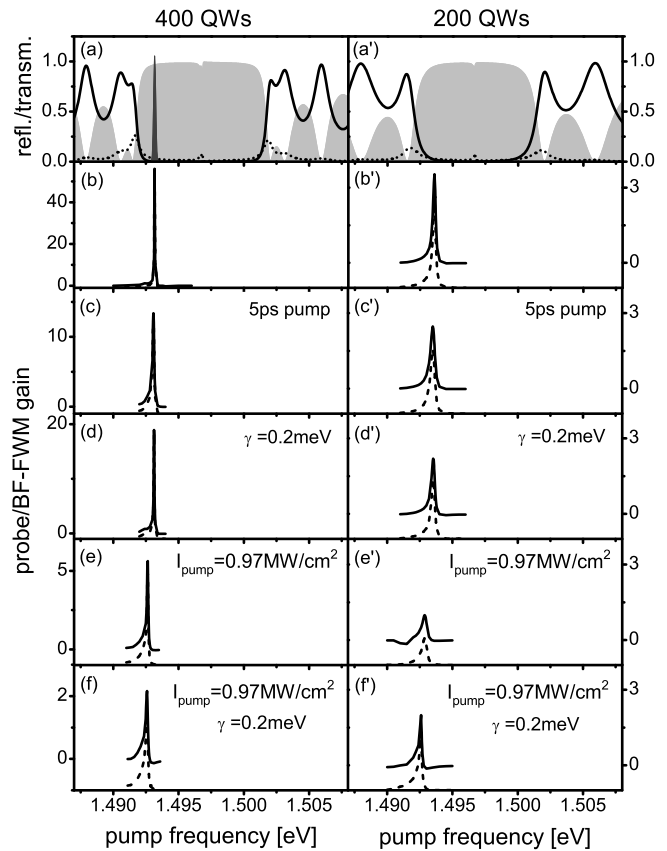}
\caption{\label{Fig2gain} Four-wave mixing induced gain in
Bragg-spaced quantum wells in a typical pump-probe setup (air on
both sides of the sample). Left column: $400$ QWs, right column:
$200$ QWs. Results are shown for co-circular excitation with
Gaussian pump and probe pulses of $10\,\mathrm{ps}$ (FWHM) and
$1\,\mathrm{ps}$ (FWHM) length, respectively. The pump-probe delay
time is zero. The pump peak intensity is
$I_{\text{pump}}\approx2.49\,\mathrm{MWcm^{-2}}$, and the exciton
dephasing is $\gamma=0.1\,\mathrm{meV}$. All other material
parameters are summarized in Ref.~\onlinecite{BSInstfootnote1}.
Deviations from these parameters are noted in each panel. (a), (a')
show the linear reflection (gray-shaded area), transmission (solid
line), and absorption (dashed line) of the samples. (a) exemplarily
includes the spectral shape of the pump pulse (dark-gray shaded
area) for the pump frequency where the maximum gain is observed in
(b). (b)-(f), (b')-(f') show the gain in the probe (solid) and
background-free four-wave mixing (dashed) directions vs. the central
pump frequency (same as central probe frequency). Note the different
scales of the vertical axes (gain axes) in the left column.}
\end{figure}

\textsc{Numerical Results \& Discussion} -- In the framework of this
theoretical approach we solve the coupled equations of motion for
the optically induced probe, BF-FWM, and pump polarizations in each
QW. The dynamics of these polarizations is calculated
self-consistently together with the time-dependent propagation of
the optical fields inside the structure. Via this self-consistent
solution of exciton-polarization dynamics and Maxwell's equations,
the exciton-exciton interactions which are local in each QW lead to
the effective polariton-polariton interaction in our theory.

From these simulations we obtain the results shown in
Fig.~\ref{Fig2gain} for pump-probe excitation of two BSQW structures
of different lengths, containing $200$ and $400$ QWs, respectively.
The linear optical properties are depicted in panels (a) and (a').
Panels (b)-(f) and (b')-(f') show the probe and BF-FWM gain,
$(F_{\text{refl.}}^{\text{probe,BF-FWM}}+F_{\text{transm.}}^{\text{probe,BF-FWM}})/F_{\text{inc.}}^{\text{probe}}-1$,
for different pump frequencies around the lower stop-band edge and
for different excitation and structural parameters as noted in each
subplot and given in the figure caption.
$F_{\text{refl.,transm.,inc.}}^{\text{probe,BF-FWM}}$ is the
time-integrated reflected, transmitted, or incoming probe or BF-FWM
intensity, respectively.

For pump excitation close to the effective polariton resonances near
the lower stop-band edge, for the 400-QW structure almost two orders
of magnitude of spectrally integrated probe and FWM gain are found
in Fig.~\ref{Fig2gain}(b). This gain is strongly peaked at a
specific pump frequency where excitation with the spectrally narrow
pump allows for resonant pairwise scattering of pump polaritons on
the polariton dispersion (blue-shifted by the repulsive
polariton-polariton interaction) of probe and BF-FWM. Note that, in
Fig.~\ref{Fig2gain}(b) this peak in the gain is spectrally narrower
than the pump spectrum itself. Additional calculations, not
represented in Fig.~\ref{Fig2gain}, show that the PCO-induced gain
is sufficient to overcome intrinsic and radiative losses of the QW
polarizations so that the unstable regime is reached. Once the
system is unstable, the probe and BF-FWM gain grow almost
exponentially with the pump length as long as the scattering into
the probe and BF-FWM beams does not significantly deplete the pump.
Note that strict exponential growth is only expected in a
steady-state analysis, not for excitation with finite lengths
pulses. Therefore in this unstable regime the gain is strongly
reduced by reducing the pump length from $10\,\mathrm{ps}$ [panel
(b)] to $5\,\mathrm{ps}$ [panel (c)]. Furthermore it is reduced by
choosing less favorable parameters, i.e., a larger intrinsic
dephasing and a lower pump intensity, as is the case in panels
(c)-(f). As shown in panels (b')-(f') of Fig.~\ref{Fig2gain}, for
the shorter 200-QW structure only moderate amounts of gain are
found. Being further away from the limit of the infinite structure,
the effective polariton resonances in the 200 QWs are not
sufficiently pronounced to reach the unstable regime for any of the
cases shown in Fig.~\ref{Fig2gain}.

\begin{figure}[t]
\includegraphics[scale=0.77]{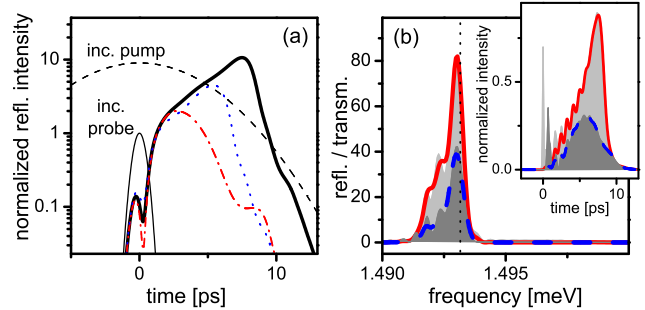}
\caption{\label{Fig3instability} (color online) (a) Time-resolved
reflected probe intensity (normalized to the incoming probe's peak
intensity), corresponding to the data shown in
Fig.~\ref{Fig2gain}(b). Results are shown for pump (thin dashed
line) and probe (thin solid line) central frequencies of
$1.49312\,\mathrm{eV}$ (dashed-dotted), $1.49316\,\mathrm{eV}$
(solid), $1.49320\,\mathrm{eV}$ (dotted). (b) Spectral domain
results for the probe (light-gray shaded area) and BF-FWM (solid
line) reflection and probe (dark-gray shaded area) and BF-FWM
(dashed line) transmission for the optimum pump frequency
$1.49316\,\mathrm{eV}$ (indicated by the dotted line). Calculations
done for a spectrally broad $200\,\mathrm{fs}$ probe. The inset
shows the corresponding time-resolved reflected and transmitted
probe and BF-FWM intensities normalized to the incoming probe's peak
intensity. In the unstable regime shown, probe and BF-FWM play equal
roles in the wave-mixing process dominating the system dynamics.
Therefore, results for these two directions are almost
indistinguishable in the plots (except for the initial probe
transmission and reflection visible in the inset).}
\end{figure}

Although the observation of large optical gain that increases with
increasing pump length [cf. panels (c), (b) of Fig.~\ref{Fig2gain}]
is indicative of the proposed instability in the BSQW structure, it
does not necessarily imply exponential growth (i.e., instability) in
the time domain. Therefore, we supplement the gain analysis in
Fig.~\ref{Fig2gain} by the time domain results shown in
Fig.~\ref{Fig3instability}(a). These results correspond to the data
in Fig.~\ref{Fig2gain}(b) around the pump frequency where the
maximum gain is found. Only the probe reflection is shown since
probe transmission and BF-FWM transmission and reflection show the
same qualitative time dependence. Results are shown for three
different pump frequencies to demonstrate the sensitivity of the
time dynamics to this parameter. Close to the optimum pump frequency
the time-resolved probe reflection shows (almost) exponential growth
over a certain time period as long as the pump pulse is strong
enough to keep the system dynamics in the unstable regime [clearly
visible for the thick solid line in Fig.~\ref{Fig3instability}(a)
between $2$ and $8\,\mathrm{ps}$]. Away from the optimum pump
frequency the exponential growth generally has a smaller growth rate
and the growth period is shorter (dotted line) or the exponential
growth is absent where the unstable regime is not reached
(dashed-dotted line). Note that, even in the unstable regime
mono-exponential growth over time is not necessarily expected;
several growing and decaying modes with different time dependencies
may contribute to the total signal. In Fig.~\ref{Fig3instability}(b)
we show spectral domain results in the unstable regime for the
optimum pump frequency. Large spectral gain for probe and BF-FWM
transmission and reflection is found. Its maximum is found close to
the pump frequency. From this we conclude that the dominant
contribution stems from (almost) degenerate FWM processes
exclusively involving polaritons on the LPB.

The above-investigated instability in the BSQW is enabled by
pump-excitation of the coherent polariton states at the lower
stop-band edge. By the strong radiative coupling of the QWs in the
structure these coherent states are spectrally well below (several
meV) the exciton resonance of the single QW. This way two-exciton
correlations that weaken the effective polariton-polariton
scattering processes driving the instability and EID losses due to
scattering off coherent excitons are strongly suppressed. Similar
effects were previously pointed out for excitation of microcavity
polaritons on the LPB.\cite{Kwong2001a,Kwong2001b,Savasta2003} In
principle the accumulation of incoherent excitons in the QWs over
time can lead to additional EID in the system which might cap the
maximal achievable amount of gain with increasing pump length.
However, as in the amplification of microcavity polaritons, for
excitation far below the exciton resonance generation of incoherent
excitons is expected to be strongly suppressed.

As noted above, the instability discussed in this paper is driven by
(almost) degenerate FWM processes. In a one-dimensional non-resonant
photonic crystal non-degenerate FWM involving photons on different
branches of the dispersion has been used\cite{Becker2006} to achieve
efficient frequency-conversion. However, for the BSQW an instability
relying on inter-branch scattering of polaritons may well be
inhibited by excessive EID in the UPB.

\textsc{Linear Stability analysis} -- Guided by the numerical
results discussed above which -- in the framework of our theoretical
approach -- include the full nonlinear polarization and field
dynamics, we have also performed a linear stability analysis for
monochromatic steady-state pump excitation. With the transfer matrix
which relates incoming and outgoing field amplitudes in the probe
and BF-FWM directions at both ends of the multiple quantum-well
structure, a temporal instability can be identified: at threshold
even for zero-incoming boundary conditions non-trivial solutions for
the outgoing fields exist. A similar analysis is used for example in
Refs.~\onlinecite{Yariv1977,Firth1990}. The following general trends
could be confirmed by the linear stability analysis: Above a certain
pump-induced threshold exciton density, instability is found in a
small spectral window at the effective (Hartree-Fock blue-shifted)
lower stop-band edge. Increasing the structure length and/or
decreasing the intrinsic dephasing $\gamma$ of the QW polarization
generally lowers the required threshold density in agreement with
the time-domain results.

\textsc{Conclusions \& Remarks} -- Based on a microscopic
many-particle theory we predict FWM instabilities in a typical
pump-probe setup in Bragg-spaced multiple quantum-well structures.
In contrast to the well known FWM instability in planar
microcavities these instabilities are not enabled by a specific
in-plane polariton dispersion but by the polariton dispersion for
propagation quasi-perpendicular to the QW planes.

We note that inclusion of semiconductor to air transitions at the
ends of the structure generally lowers the threshold because it
effectively increases the coupling strength of the QW polarizations
to the propagating fields. However, this ``cavity enhancement'' is
not required for the reported instability.  Further research, e.g.,
into polarization dependencies would be desirable.

\textsc{Acknowledgments} -- The authors are indebted to A. L. Smirl,
University of Iowa, for many enlightening discussions. This work has
been supported by ONR, DARPA, JSOP. \mbox{S. Schumacher} gratefully
acknowledges support by the Deutsche Forschungsgemeinschaft (DFG,
project No. SCHU~1980/3-1).

\bibliography{../../../literature}

\begin{thebibliography}{10}
\bibitem{BSInstfootnote1}
Parameters used are, the background refractive index $n_{\text{bg}}=3.61$,
  effective electron mass $m_e=0.067\,m_0$, and effective hole mass
  $m_h=0.1\,m_0$. The corresponding exciton binding energy is
  $E_b^x\approx13\,\mathrm{meV}$ and the bulk Bohr-radius is
  $a_0^x\approx170\,\AA$. For the evaluation of the theory also the 1s exciton
  energy $\varepsilon_x=1.4965\,\mathrm{eV}$, the dipole coupling
  $d_{cv}=4\,e\AA$, and dephasing $\gamma=0.1\,\mathrm{meV}$ are required. The
  Bragg energy used is
  $\varepsilon_{\text{Bragg}}=\varepsilon_x+0.5\,\mathrm{meV}=1.497\,\mathrm{e%
V}$.
\end{thebibliography}
\bibliographystyle{prsty}

\end{document}